\documentclass[aip,amsmath,amssymb,reprint]{revtex4-2}
\usepackage{graphicx}
\usepackage{hyperref}
\graphicspath{{figs/}}
\usepackage{dcolumn}
\usepackage{bm}
\usepackage[utf8]{inputenc}
\usepackage[T1]{fontenc}
\usepackage{mathptmx}
\usepackage{etoolbox}
\usepackage{color}
\newcommand{\NEU}[1]{ #1 }
\makeatletter
\def\@email#1#2{%
 \endgroup
 \patchcmd{\titleblock@produce}
  {\frontmatter@RRAPformat}
  {\frontmatter@RRAPformat{\produce@RRAP{*#1\href{mailto:#2}{#2}}}\frontmatter@RRAPformat}
  {}{}
}%
\makeatother

\begin{document}

%%%%%%%%%%%%%%%%%%%%%%%%%%%%%%%%%%%%%%%%%
\title{Rheology of dilute granular gas mixtures where the grains interact via a square shoulder and well potential}

%\author{Kiwamu Yoshii (吉井究)}
\author{Kiwamu Yoshii}
%\email{k\_yoshii@fm.me.es.osaka-u.ac.jp}
\affiliation{Department of Mechanical Science and Bioengineering, Osaka University, 1--3 Machikaneyama, Toyonaka, Osaka 560--8531, Japan}

%\author{Satoshi Takada (髙田智史)}
\author{Satoshi Takada}
%\email{takada@go.tuat.ac.jp}
\affiliation{Department of Mechanical Systems Engineering and Institute of Engineering, Tokyo University of Agriculture and Technology, 2--24--16, Naka-cho, Koganei, Tokyo 184--8588, Japan}

%\author{Kosuke Kurosawa (黒澤耕介)}
\author{Kosuke Kurosawa}
%\email{kosuke.kurosawa@perc.it-chiba.ac.jp}
\affiliation{Planetary Exploration Research Center, Chiba Institute of Technology, 2--17--1, Narashino, Tsudanuma, Chiba 275--0016, Japan}

\author{Thorsten P\"{o}schel\thanks{corresponding author}}
\email{thorsten.poeschel@fau.de}
\affiliation{
Lehrstuhl f\"{u}r Multiscale Simulation, Friedrich-Alexander-Universit\"{a}t Erlangen-N\"{u}rnberg, Cauerstra{\ss}e 3, 91058 Erlangen, Germany}

\date{\today}

%%%%%%%%%%%%%%%%%%%%%%%%%%%%%%%%%%%%%%%%%
\begin{abstract}
  We develop the rheology of a dilute granular gas mixture. Motivated by the interaction of charged granular particles, we assume that the grains interact via a square shoulder and well potential. Employing kinetic theory, we compute the temperature and the shear viscosity as functions of the shear rate. Numerical simulations confirm our results above the critical shear rate. At a shear rate below a critical value, clustering of the particles occurs.
\end{abstract}
\maketitle

%%%%%%%%%%%%%%%%%%%%%%%%%%%%%%%%%%%%%%%%%
\section{Introduction}
In many situations of practical importance, the particles in granular gases are electrically charged, either  in the course of collisions through the effect of triboelectricity or in technical applications, e.g., the toner in copy machines \cite{schein2013electrophotography}. Although charges substantially influence on the dynamics of granular gases, almost all contributions to the kinetic theory of granular gases neglect charges. That is, charged granular gases have been considered theoretically and numerically in only a few studies in kinetic theory, e.g. \cite{scheffler2002collision,poschel2003long,takada2017homogeneous,takada2022transport,singh2018early,singh2019electrification}. Using the concept of a modified coefficient of restitution, the traditional kinetic theory of granular gases was employed to model gases of granular particles of identical charge\cite{poschel2003long,takada2017homogeneous,takada2022transport}. It was found that the freely cooling state deviates from Haff's law\cite{haff1983grain} at the later stage of the gas' evolution, that is, the granular temperature decays logarithmically slow\cite{scheffler2002collision,poschel2003long,takada2017homogeneous}. In a certain transition state (characterized by the transition velocity), the velocity distribution of the gas deviates significantly from the Gaussian distribution\cite{takada2017homogeneous}, leading to a nontrivial behavior of the transport coefficients\cite{takada2022transport}.

The model used in Ref. [\onlinecite{takada2022transport}] neglects, however, the collision geometry, particularly, the collision angle dependence of the coefficient of restitution. Moreover, it is assumed that all particles carry the same charge. The situation becomes significantly more complicated in the case of positively and negatively charged granular particles. Here, the theory is similar to polydisperse mixtures, which have been studied intensively in the past twenty years\cite{mcnamara1998energy,garzo1999homogeneous,dahl2002kinetic,garzo2002tracer,montanero2002rheological,montanero2003energy,garzo2003effect,brilliantov2004kinetic,alam2005energy,garzo2007enskog,garzo2007enskog2,murray2012enskog,garzo2019granular}. The most significant feature of granular mixtures is the violation of the energy equipartition\cite{mcnamara1998energy}, which inspired the concept of a partial temperature for each species. These temperatures depend on the particle mass and size and the gas density of the corresponding species. For our system of particles carrying different charges, the partial temperature ratios depend, moreover, on the type of the interaction between the particles. In the kinetic theory, the partial temperatures are determined self-consistently to satisfy the energy balance equations.

In the current paper, we consider the interaction of particles characterized by square shoulder and square well potentials\cite{bannerman2010exact}. This allows for the analytical description of the collision process and a semi-analytical computation of the transport coefficients. We compare the results to earlier findings for gases whose particles interact via a square well potential\cite{takada2016kinetic,takada2018rheology}.

In the next section, we introduce our model and consider the corresponding scatter process. In Sec. \ref{sec:results}, we consider the kinetic theory under a plane shear and derive the shear viscosity and the shear rate as functions of the temperature. Finally, in Sec. \ref{sec:simulation}, we verify the results through numerical molecular dynamics simulations. Appendix \ref{sec:Lambda} details the calculation of the collision integral. 
\NEU{Appendix \ref{sec:evolution} shows how the temperature and the viscosity converge to the mean values against the number of collisions.}

%%%%%%%%%%%%%%%%%%%%%%%%%%%%%%%%%%%%%%%%%
\section{Particle Interaction Model and Scatter process}
\subsection{Model}
\begin{figure}[htbp]
	\includegraphics[width=\linewidth]{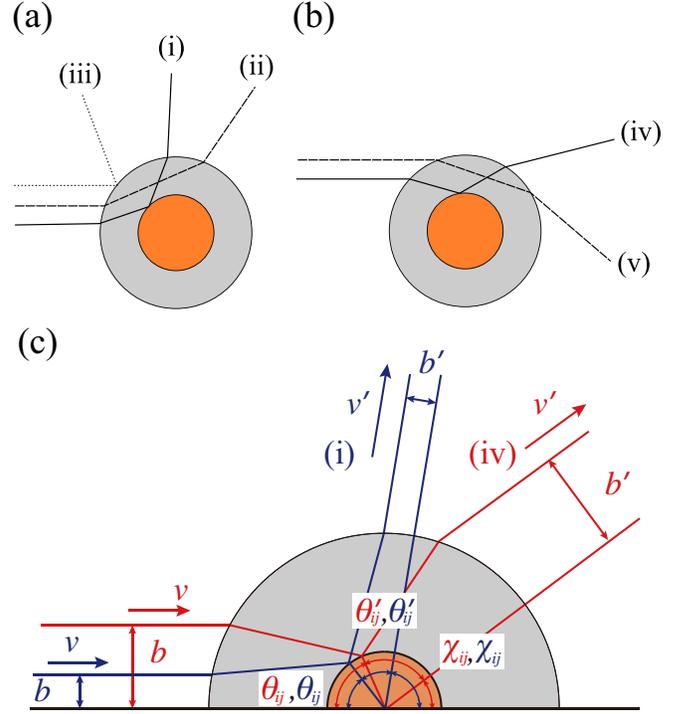}
	\caption{(a) Sketch of the collision of particles carrying equal charges: (i) inelastic hard-core collision at $r=\NEU{d_{ij}}$ (solid line), (ii) elastic grazing collision (dashed line), (iii) elastic hard-core collision at $r=\lambda \NEU{d_{ij}}$ (dotted line). (b) The same for differently charged particles: (iv) inelastic hard-core collision at $r=\NEU{d_{ij}}$, and (v) elastic grazing collision (dashed line). The shaded regions illustrate the (a) shoulder and (b) well regions ($\NEU{ d_{ij}}<r\le \lambda \NEU{ d_{ij}}$).
\NEU{ (c) Schematic picture of inelastic hard-core collision processes. 
    The blue (red) line represents the trajectory of a particle when it interacts with an identical (a different) particle.
    }}
	\label{fig:collision}
\end{figure}

We consider a dilute gas of granular particles of different species, $i$, characterized by their masses, $m_i$, and diameter, $d_i$, interacting pairwise via the potential 
\begin{equation}
	U_{ij}(r)=
	\begin{cases}
		\infty & (r\le d_{ij})\\
		\pm \varepsilon & (d_{ij}<r\le \lambda d_{ij})\\
		0 & (r>\lambda d_{ij})\,,
	\end{cases}
\end{equation}
where $d_{ij}\equiv (d_i+d_j)/2$, $r$ is the distance of the particles, $\varepsilon$ characterizes the strength of the repulsive force, and $\lambda$ is the shoulder width ratio.
\begin{figure}[htbp]
  \includegraphics[width=0.9\linewidth]{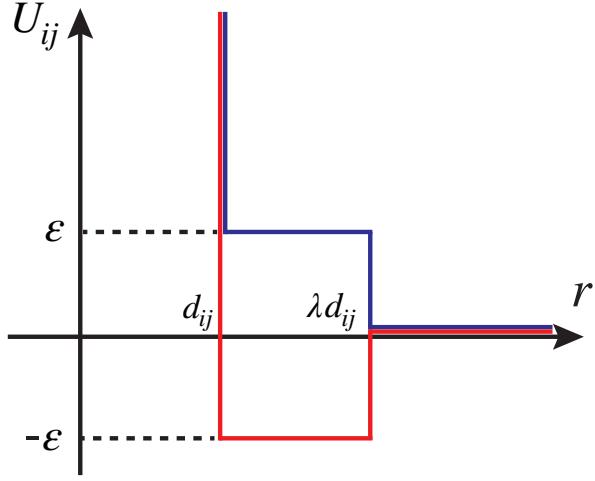}
  \caption{\NEU{Sketch of the potential between particles.  There exist square shoulder and well regions in $d_{ij}\le r \le \lambda d_{ij}$. The blue (red) line represents the interaction of identical (different) particles.}}
  \label{fig:potential}
\end{figure}
The plus sign stands for the interaction of identical particles, and the minus sign stands for particles of different species. 
\NEU{The potential between particles is shown in Fig. \ref{fig:potential}.}
As long as particles do not touch one another, $r>d_{ij}$, the energy of the relative motion is conserved, and the particles are accelerated or deaccelerated in the shoulder and well regions. If the particles touch each other at $r=d_{ij}$, an inelastic collision occurs, characterized by the coefficient of restitution, $e$. For small dissipation, considered here, $e\lesssim 1$. Figure \ref{fig:collision} illustrates the interaction of the particles.

%%%%%%%%%%%%%%%%%%%%%%%%%%%%%%%%%%%%%%%%%
%%%%%%%%%%%%%%%%%%%%%%%%%%%%%%%%%%%%%%%%%

%%%%%%%%%%%%%%%%%%%%%%%%%%%%%%%%%%%%%%%%%
\subsection{Scatter processes}
\begin{figure*}[htbp]
	\includegraphics[width=0.8\linewidth]{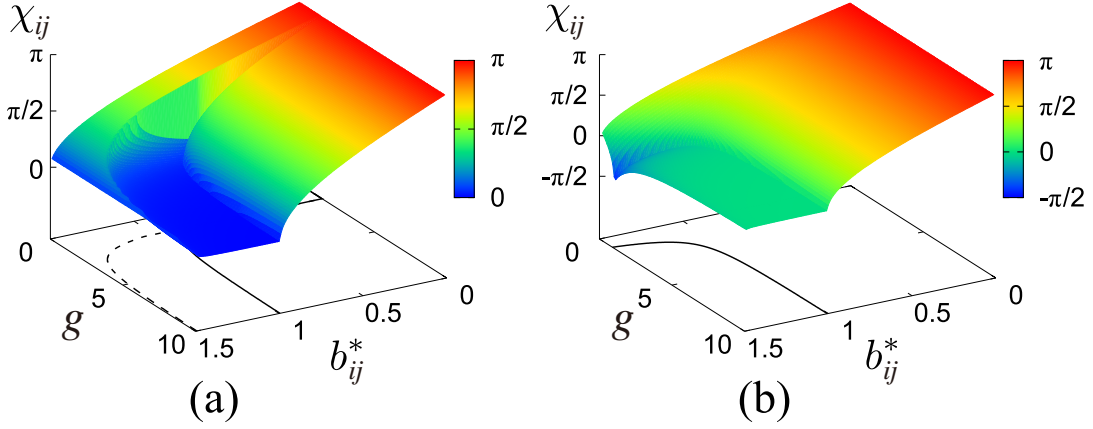}
	\caption{The scatter angle, \NEU{$\chi_{ij}(b,v)$}, as a function of the \NEU{dimensionless} collision parameter \NEU{$b^*_{ij}$} and the \NEU{dimensionless} relative velocity \NEU{$g$} between the particles of (a) the same species, and (b) different species.
	The color represents the magnitude of the scatter angle.}
	\label{fig:chi_b_v}
\end{figure*}

We consider scattering for repulsive and attractive potentials. When two particles of species $i$ and $j$ collide, the angle \NEU{$\theta_{ij}$} between the incidental asymptote and the closest approach is given by\cite{goldstein2002classical}
\begin{equation}
	\theta_{ij} = b\int_0^{u_0} \frac{du}{\displaystyle \sqrt{1-b^2u^2 -\frac{\NEU{2}}{\NEU{m_{ij}}v^2}U_{ij}\left(\frac{1}{u}\right)}}\,,
	\label{eq:theta_def}
\end{equation}
with $m_{ij}\equiv m_im_j/(m_i+m_j)$, \NEU{$u \equiv 1/r$,}  and the refractive index\cite{landau1976mechanics,goldstein2002classical} 
\begin{align}
    \nu_{\rm r}(v)
    =
    \begin{cases}
	\nu_{\rm r}^{\rm rep}=
	\begin{cases}
		0 & (m_{ij}v^2 \le 2\varepsilon)\\
		\displaystyle \sqrt{1-\frac{2\varepsilon}{m_{ij}v^2}} &(m_{ij}v^2>2\varepsilon)
	\end{cases}
	&({\rm repulsive}) \\[25pt]
	\displaystyle \nu_{\rm r}^{\rm att}=
	\sqrt{1+\frac{2\varepsilon}{m_{ij}v^2}}  & ({\rm attractive})
    \end{cases},
\end{align}
\NEU{where $b$ is} the impact parameter and \NEU{$v$ is the relative velocity.
Depending on these parameters}, there are three types of collisions illustrated in Fig. \ref{fig:collision}\NEU{(a),(b).
In Fig. \ref{fig:collision}(c), we exemplify of inelastic hard-core collision processes.
}
And $u_0$ is determined by the collision process.
If $\min(1, \nu_{\rm r})\lambda d_{ij}<b\le \lambda d_{ij}$ (which occurs only for repulsive interaction), the particle cannot enter the shoulder region, which means that the reflection at $r=\lambda d_{ij}$ occurs.
In this case, Eq. \eqref{eq:theta_def} becomes
\begin{equation}
	\theta_{ij} = b\int_0^{1/(\lambda d_{ij})} \frac{du}{\sqrt{1-b^2u^2}} 
	= \sin^{-1} \frac{b}{\lambda d_{ij}}
\end{equation}
and the scatter angle is
\begin{equation}
  \chi_{ij} = \pi - 2\theta_{ij} = \pi - 2\sin^{-1} \frac{b}{\lambda d_{ij}}.
  \label{eq:chi_3}
\end{equation}
On the other hand, when $b\le \min(1, \nu_{\rm r})\lambda d_{ij}$, the particle can enter the potential region and Eq. \eqref{eq:theta_def} becomes
\begin{align}
	\theta_{ij}
	&= b\int\limits_0^{1/(\lambda d_{ij})} \frac{du}{\sqrt{1-b^2u^2}} 
       + b\int\limits_{1/(\lambda d_{ij})}^{u_0} \frac{du}{\displaystyle\sqrt{1-b^2u^2\mp\frac{2\varepsilon}{m_{ij}v^2}}}\nonumber\\
	&= \sin^{-1} \frac{b}{\lambda d_{ij}}+ b\int_{1/(\lambda d_{ij})}^{u_0} \frac{du}{\sqrt{\nu_{\rm r}^2-b^2u^2}}.
	\label{eq:theta_def2}
\end{align} 
There are two types of collisions in this case.
First, if $\min(\nu_{\rm r}, \lambda)d_{ij}< b \le \min(1, \nu_{\rm r})\lambda d_{ij}$, the cores of the particle cannot touch, such that $u_0=\nu_{\rm r}/b$. Equation \eqref{eq:theta_def2} becomes then
\begin{align}
	\theta_{ij} 
	= \frac{\pi}{2}+\sin^{-1} \frac{b}{\lambda d_{ij}} - \sin^{-1} \frac{b}{\nu_{\rm r}\lambda d_{ij}},
\end{align}
and the scatter angle is
\begin{equation}
  \chi_{ij} = \pi - 2\theta_{ij} = 2\sin^{-1} \frac{b}{\nu_{\rm r}\lambda d_{ij}} - 2\sin^{-1} \frac{b}{\lambda d_{ij}}.
  \label{eq:chi_2}
\end{equation}
Second, if $0\le b\le \min(\nu_{\rm r},\lambda)d_{ij}$, the cores of the particles touch, thus, the collision is inelastic.
With Eq. \eqref{eq:theta_def}, we obtain
\begin{align}
	\theta_{ij} 
  = \sin^{-1} \frac{b}{\nu_{\rm r}d_{ij}}+\sin^{-1} \frac{b}{\lambda d_{ij}} - \sin^{-1} \frac{b}{\nu_{\rm r}\lambda d_{ij}}.
  \label{eq:def_theta_hard_core}
\end{align}
In this case, the angle after the collision from the closest distance $\theta^\prime_{ij}$ is different from $\theta_{ij}$ because the impact parameter $b^\prime$ and the velocity $v^\prime$ after the collision at $r\to\infty$ are also different from $b$ and $v$, respectively (\NEU{see also Fig.~\ref{fig:collision}(c)}).
Exploiting the conservation of the angular momentum, we obtain:
%\begin{widetext}
\begin{align}
	\theta_{ij}^\prime
	&= \sin^{-1} \frac{b^\prime}{\nu_{\rm r}(v^\prime) d_{ij}}+\sin^{-1} \frac{b^\prime}{\lambda d_{ij}} 
		- \sin^{-1} \frac{b^\prime}{\nu_{\rm r}(v^\prime) \lambda d_{ij}} \nonumber\\
	&= \sin^{-1} \frac{b}{\nu_{\rm r}d_{ij}}+\sin^{-1} \frac{b}{\lambda d_{ij}} - \sin^{-1} \frac{b}{\nu_{\rm r} \lambda d_{ij}}\nonumber\\
		&\hspace{1em}+ \left(\frac{b}{\sqrt{\nu_{\rm r}^2d_{ij}^2-b^2}}
		+\frac{b\nu_{\rm r}^2}{\sqrt{\lambda^2d_{ij}^2-b^2}}- \frac{b}{\sqrt{\nu_{\rm r}^2\lambda^2d_{ij}^2-b^2}}\right)\nonumber\\
		&\hspace{2em}\times\frac{1-e^2}{2}\cos^2 \Theta_{ij}, \label{eq:def_theta_prime}
\end{align}
%\end{widetext}
with 
\begin{equation}
    \Theta_{ij} \equiv \cos^{-1}\sqrt{1-\frac{b^2}{\nu_{\rm r}^2d_{ij}^2}}.
\end{equation}
With Eqs. \eqref{eq:def_theta_hard_core} and \eqref{eq:def_theta_prime}, the scatter angle \NEU{$\chi_{ij}$} for a hard-core collision is given by
\begin{equation}
  \begin{split}
    \chi_{ij} &= \pi - \theta_{ij} - \theta_{ij}^\prime\\
    &=\chi_{ij}^{(0)}+(1-e^2)\chi_{ij}^{(1)}+\mathcal{O}((1-e^2)^2), \label{eq:chi_1}
      \end{split}
\end{equation}
with
\begin{subequations}
\begin{align}
	\chi_{ij}^{(0)}&= \pi - 2\sin^{-1} \frac{b}{\nu_{\rm r}d_{ij}}-2\sin^{-1} \frac{b}{\lambda d_{ij}} 
		+ 2\sin^{-1} \frac{b}{\nu_{\rm r} \lambda d_{ij}},\\
	\chi_{ij}^{(1)}&= -\left(\frac{b}{\sqrt{\nu_{\rm r}^2d_{ij}^2-b^2}}
		+\frac{b\nu_{\rm r}^2}{\sqrt{\lambda^2d_{ij}^2-b^2}}- \frac{b}{\sqrt{\nu_{\rm r}^2\lambda^2d_{ij}^2-b^2}}\right)\nonumber\\
		&\hspace{1em}\times\frac{1}{2}\cos^2 \Theta_{ij}.
\end{align}
\end{subequations}
The above results are summarized and plotted in Table \ref{fig:collision_type} and Fig. \ref{fig:chi_b_v}, respectively.
%%%%%%%%%%%%%%%%%%%%%%%%%%%%%%%%%%%%%%%%%
\begin{table*}[htpb]
	\caption{Three types of collisions.}
 	\begin{tabular}{c|c|c|c}
 	\hline\hline
 	 & (i) hard-core & (ii) grazing & (iii) hard-core \\
 	 & (inelastic) & (elastic) & (elastic) \\ \hline
 	condition & $0\le b/d_{ij} \le \min(\nu_{\rm r},\lambda)$ & $\min(\nu_{\rm r},\lambda)<b/d_{ij}\le \min(1, \nu_{\rm r})\lambda$ 
 		& $\min(1, \nu_{\rm r}) \lambda< b/d_{ij}\le \lambda$ \\ \hline
	\NEU{$\chi_{ij}$} & Eq. \eqref{eq:chi_1} & Eq. \eqref{eq:chi_2} & Eq. \eqref{eq:chi_3} \\
 	\hline\hline
	\end{tabular}
	\label{fig:collision_type}
\end{table*}
%%%%%%%%%%%%%%%%%%%%%%%%%%%%%%%%%%%%%%%%%
%%%%%%%%%%%%%%%%%%%%%%%%%%%%%%%%%%%%%%%%%

%%%%%%%%%%%%%%%%%%%%%%%%%%%%%%%%%%%%%%%%%
\section{Boltzmann equation and derivation of the shear viscosity}
\label{sec:results}
We construct the kinetic theory for square shoulder potential based on the scatter processes, as discussed in the previous section. Let us consider the system under  homogeneous shear. 
%We can write the Boltzmann equation for species $i$ under a plane shear as
\NEU{The Boltzmann kinetic equation for species $i$ of a dilute granular system under plane shear reads \cite{chamorro2015non,hayakawa2017kinetic}}
\begin{equation}
	\left(\frac{\partial}{\partial t} -\dot\gamma V_{1y}\frac{\partial}{\partial V_{1x}}\right)f_i(\bm{V}_1,t)
	=\sum_{j=1}^2J_{ij}(\bm{V}_1|f_i, f_j),
	\label{eq:Boltzmann}
\end{equation}
where $\dot\gamma$ is the shear rate, $\bm{V}_i\equiv \bm{v}_i-\dot\gamma y_i \hat{\bm{e}}_x$ is the peculiar velocity with the unit vector in $x$-direction, $\hat{\bm{e}}_x$, \NEU{$f_i(\bm{V}, t)$} is the velocity distribution function, and $J_{ij}(\bm{V}|f_i, f_j)$ is the collision integral defined by
\begin{align}
	&J_{ij}(\bm{V}_1|f_i, f_j)\nonumber\\
	&\equiv \int d\bm{V}_2 \int d\widehat{\bm{\sigma}} |\bm{v}_{12}\cdot \widehat{\bm{\sigma}}| 
	\left[\chi_e \sigma_{\rm s}(\chi_{ij}, V_{12}^{\prime\prime}) f_i(\bm{V}_1^{\prime\prime},t)f_j(\bm{V}_2^{\prime\prime},t)\right.\nonumber\\
	&\hspace{8em}\left. - \sigma_{\rm s}(\chi_{ij},V_{12})f_i(\bm{V}_1,t)f_j(\bm{V}_2,t)\right].
\end{align}
Here, $\chi_e$ is the Jacobian of the transformation between the pre-collisional and post-collisional velocities, $\sigma_{\rm s}$ is the cross section between particles $1$ and $2$, and $\widehat{\bm{\sigma}}$ is a unit vector parallel to $\bm{r}_1-\bm{r}_2$.
The relationship between the pre-collisional and the post-collisional velocities is
\begin{equation}
  \begin{split}
%	\begin{cases}
	\displaystyle
	\bm{v}_1^\prime 
	&= \bm{v}_1 - \mu_{ji}A_{ij}\cdot (\bm{v}_{12}\cdot \widehat{\bm{\sigma}})\widehat{\bm{\sigma}}\\
	%%%
	\displaystyle
	\bm{v}_2^\prime 
	&= \bm{v}_2 + \mu_{ij}A_{ij}\cdot (\bm{v}_{12}\cdot \widehat{\bm{\sigma}})\widehat{\bm{\sigma}}
        % \end{cases}
  \end{split}
        \label{eq:relation_pre_post_vel}
\end{equation}
with $\mu_{ij}\equiv m_i/(m_i + m_j)$ and
\begin{align}
	A_{ij} =
	\begin{cases}
	\displaystyle 1+\sqrt{1-(1-e^2)\nu_{\rm r}^2\frac{\cos^2\Theta_{ij}}{\cos^2\theta_{ij}}}
	& (b<\min(\nu_{\rm r},\lambda)d_{ij})\\
	2 & ({\rm otherwise})
	\end{cases}.
\end{align}

Multiplying Eq. \eqref{eq:Boltzmann} with $m_i v_\alpha v_\beta$ and integrating over $\bm{v}$, the evolution of the stress tensor \NEU{for species $i$} can be obtained as
\begin{equation}
	\partial_t P^{(i)}_{\alpha\beta} 
	+ \dot\gamma \left(\delta_{\alpha x}P^{(i)}_{y\beta} + \delta_{\beta x}P^{(i)}_{\alpha y} \right)
	= -\sum_{j=1}^2\Lambda^{(ij)}_{\alpha \beta}\label{eq:partial_t_P}
\end{equation}
with the kinetic part of the partial stress tensor $P^{(i)}_{\alpha \beta}\equiv \int d\bm{v} \NEU{m_i} V_\alpha V_\beta f_i(\bm{V},t)$, \NEU{and $\delta_{\alpha \beta}$ is the Kronecker's delta function.}
Here, $\Lambda^{(ij)}_{\alpha \beta}$ is the second moment of the collision integral defined by
\begin{equation}
	\Lambda^{(ij)}_{\alpha \beta} 
	\equiv -\int d\bm{v}_1 m_i V_{1,\alpha} V_{1, \beta} J_{ij}(\bm{V}_1|f_i, f_j).
        \label{eq:Lambda_def}
\end{equation}
We employ Grad's moment method to write the approximate velocity distribution function\cite{montanero2002rheological,montanero2003energy,garzo2003effect}:
\begin{equation}
	f_i(\bm{V}) = f_{i,{\rm M}}(\bm{V})\left[1+\frac{m_i}{2T_i}\Pi^{(i)}_{\alpha\beta}\left(V_\alpha V_\beta - \frac{1}{3}V^2 \delta_{\alpha\beta}\right)\right],
	\label{eq:Grad}
\end{equation}
where
\begin{equation}
	\Pi^{(i)}_{\alpha\beta}=\frac{P^{(i)}_{\alpha\beta}}{p_i}-\delta_{\alpha\beta},
\end{equation}
with the partial pressure $p_i = n_i T_i$, and $f_{i, {\rm M}}(\bm{V})$ is the Maxwell distribution function for species $i$ given by
\begin{equation}
	f_{i, {\rm M}}(\bm{V}) = n_i\left(\frac{m_i}{2\pi T_i}\right)^{3/2}\exp\left(-\frac{m_iV^2}{2T_i}\right).
\end{equation}
Here, we introduce the \NEU{number density and the} partial temperature \NEU{of species $i$, $n_i$ and} $T_i$, respectively.
\NEU{The latter} is, in general, different from the mean temperature, $T$.
Once we can obtain the partial temperatures, $T_1$ and $T_2$, the mean temperature can be written as $nT = n_1T_1 + n_1T_2$, \NEU{with the total number density $n = n_1 + n_2$ \cite{garzo1999homogeneous}.}

Substituting Eq. \eqref{eq:Grad} into Eq. \eqref{eq:Lambda_def}, a long calculation \cite{takada2018rheology,montanero2002rheological,montanero2003energy,garzo2003effect} delivers the tensor $\overleftrightarrow{\Lambda}^{(ij)}$:
\begin{equation}
    \Lambda^{(ij)}_{\alpha \beta}
    = \zeta_{ij} p \delta_{\alpha\beta}
    + \nu^{(1)}_{ij} \left(P^{(i)}_{\alpha\beta}-p_i \delta_{\alpha\beta}\right)
	+ \nu^{(2)}_{ij} \left(P^{(j)}_{\alpha\beta}-p_i \delta_{\alpha\beta}\right),
    \label{eq:Lambda_ij}
\end{equation}
where $p=nT$ is the mean static pressure and the definitions
% Here, $\zeta_{ij}$, $\nu^{(1)}_{ij}$, and $\nu^{(2)}_{ij}$ are, respectively, defined by
\cite{takada2018rheology}
\begin{subequations}\label{eq:def_zeta_nu}
\begin{align}
	\zeta_{ij}
	&\equiv \frac{16\sqrt{\pi}}{3}x_i x_j n d_{ij}^2 \frac{m_{ij}}{\overline{m}} v_{\rm T}
		\left(\frac{\vartheta_i \vartheta_j}{\vartheta_i+\vartheta_j}\right)^{3/2}\nonumber\\
		&\hspace{1em}\times\int_0^\infty dg \int_0^\infty db_{ij}^* 
		A_{ij} b_{ij}^* g^5\sin^2\frac{\chi_{ij}}{2}\nonumber\\
		&\hspace{1em}\times
		\left(\frac{2\vartheta_j}{\vartheta_i+\vartheta_j}-\mu_{ji}A_{ij}\right)
		\exp\left(-\frac{\vartheta_i \vartheta_j}{\vartheta_i+\vartheta_j} g^2\right),
	\label{eq:zeta_ij}\\
	%%%%%%
	\nu^{(1)}_{ij}
	&\equiv \frac{32\sqrt{\pi}}{3}\mu_{ji} n_j d_{ij}^2 v_{\rm T}\vartheta_i
		\left(\frac{\vartheta_i \vartheta_j}{\vartheta_i+\vartheta_j}\right)^{5/2}\nonumber\\
		&\hspace{1em}\times\int_0^\infty dg \int_0^\infty db_{ij}^* 
		A_{ij} b_{ij}^* g^5\sin^2\frac{\chi_{ij}}{2} 
		\exp\left(-\frac{\vartheta_i \vartheta_j}{\vartheta_i+\vartheta_j} g^2\right)\nonumber\\
		&\hspace{1em}\times\left\{1+\frac{\vartheta_j}{5}g^2\left[\frac{2\vartheta_j}{\vartheta_i+\vartheta_j} 
				-\mu_{ji}A_{ij}\left(1-\frac{3}{2}\cos^2\frac{\chi_{ij}}{2}\right)\right]\right\},
	\label{eq:nu_ij_i}\\
	%%%%%%
	\nu^{(2)}_{ij}
	&\equiv \frac{32\sqrt{\pi}}{3}\mu_{ij} n_i d_{ij}^2 v_{\rm T}\vartheta_j
		\left(\frac{\vartheta_i \vartheta_j}{\vartheta_i+\vartheta_j}\right)^{5/2}\nonumber\\
		&\hspace{1em}\times\int_0^\infty dg \int_0^\infty db_{ij}^* 
		A_{ij} b_{ij}^* g^5\sin^2\frac{\chi_{ij}}{2} 
		\exp\left(-\frac{\vartheta_i \vartheta_j}{\vartheta_i+\vartheta_j} g^2\right)\nonumber\\
		&\hspace{1em}\times\left\{-1+\frac{\vartheta_j}{5}g^2\left[\frac{2\vartheta_i}{\vartheta_i+\vartheta_j} 
				-\mu_{ji}A_{ij}\left(1-\frac{3}{2}\cos^2\frac{\chi_{ij}}{2}\right)\right]\right\},\label{eq:nu_ij_j}
 \end{align}
\end{subequations}
where we have introduced the fraction of species $i$ as $x_i\equiv n_i/n$, the thermal velocity $v_{\rm T}\equiv \sqrt{2T/\overline{m}}$, the dimensionless collision parameter $\NEU{b_{ij}^* \equiv b/d_{ij}}$, $\vartheta_i\equiv m_i T/(\overline{m} T_i)$, and the dimensionless relative velocity $g \equiv v_{12}/v_{\rm T}$ with $\overline{m}\equiv(m_1+m_2)/2$.
These results are consistent with the previous studies \cite{montanero2002rheological,montanero2003energy,garzo2003effect,garzo2002tracer} in the hard-core limit.
The analytical solutions of  Eqs. \eqref{eq:zeta_ij}--\eqref{eq:nu_ij_j} are not known, therefore, we have to rely on numerical evaluation.
It is also noted that the control parameter is always the shear rate in simulations or experiments, while the temperature determines all quantities in the treatment of the kinetic theory.
In the following, we write all quantities as functions of the temperature.

Let us solve Eq. \eqref{eq:partial_t_P} in the steady state.
With the aid of Eq. \eqref{eq:Lambda_ij}, we can obtain a set of the equations:
\begin{subequations}
\begin{align}
	&\dot\gamma P^{(1)}_{yy} 
	= -\left(\nu^{(1)}_{11} +\nu^{(1)}_{12} + \nu^{(2)}_{11}\right)P^{(1)}_{xy} -\nu^{(2)}_{12} P^{(2)}_{xy},\label{eq:eq1}\\
	%%%
	&\dot\gamma P^{(2)}_{yy} 
	= -\nu^{(2)}_{21} P^{(1)}_{xy} -\left(\nu^{(1)}_{21} +\nu^{(1)}_{22}+\nu^{(2)}_{22}\right)P^{(2)}_{xy},\label{eq:eq2}\\
	&0 = -\left(\zeta_{11}+\zeta_{12}\right)p -\left(\nu^{(1)}_{11} +\nu^{(1)}_{12} + \nu^{(2)}_{11}\right)
		\left(P^{(1)}_{yy}-p_1\right) \nonumber\\
		&\hspace{2em}- \nu^{(2)}_{12}\left(P^{(2)}_{yy}-p_2\right),\label{eq:eq3}\\
	&0 = -\left(\zeta_{21}+\zeta_{22}\right)p - \nu^{(1)}_{21}\left(P^{(1)}_{yy}-p_1\right)\nonumber\\
		&\hspace{2em}-\left(\nu^{(1)}_{21} +\nu^{(1)}_{22}+\nu^{(2)}_{22}\right)\left(P^{(2)}_{yy}-p_2\right),\label{eq:eq4}\\
	&2\dot\gamma P^{(1)}_{xy} = -3\left(\zeta_{11}+\zeta_{12}\right)p,\label{eq:eq5}\\
	&2\dot\gamma P^{(2)}_{xy} = -3\left(\zeta_{21}+\zeta_{22}\right)p.\label{eq:eq6}
\end{align}
\label{eqs25}
\end{subequations}
To solve Eqs. \ref{eqs25} simultaneously, first, we note that Eqs. \eqref{eq:eq3} and \eqref{eq:eq4}, deliver the partial stresses:
%$P^{(i)}_{yy}$ is given by
\begin{subequations}
\begin{align}
	P^{(1)}_{yy}&= p_1 + p\frac{\mathcal{N}_1}{\mathcal{D}},\label{eq:P1yy} \\
	P^{(2)}_{yy}&= p_2 + p\frac{\mathcal{N}_2}{\mathcal{D}},\label{eq:P2yy}
\end{align}
\end{subequations}
with
\begin{subequations}
\begin{align}
	\mathcal{N}_1
	&\equiv -(\zeta_{11}+\zeta_{12})\left(\nu^{(1)}_{21} +\nu^{(1)}_{22}+\nu^{(2)}_{22}\right)
	+(\zeta_{21}+\zeta_{22})\nu^{(2)}_{12},\\
	\mathcal{N}_2
	&\equiv -(\zeta_{21}+\zeta_{22})\left(\nu^{(1)}_{11} +\nu^{(1)}_{12} + \nu^{(2)}_{11}\right)
	+(\zeta_{11}+\zeta_{12})\nu^{(2)}_{21},\\
	\mathcal{D}
	&\equiv \left(\nu^{(1)}_{11} +\nu^{(1)}_{12} + \nu^{(2)}_{11}\right)\left(\nu^{(1)}_{21} +\nu^{(1)}_{22}+\nu^{(2)}_{22}\right)
	-\nu^{(2)}_{12}\nu^{(2)}_{21}.
\end{align}
\end{subequations}
For convenience, we introduce $\eta_i \equiv -P^{(i)}_{xy}/\dot\gamma$. 
Using this definition, from Eqs. \eqref{eq:eq1}, \eqref{eq:eq2}, \eqref{eq:P1yy}, and \eqref{eq:P2yy} we obtain
\begin{subequations}
\begin{align}
	\eta_1
	&=\frac{P^{(1)}_{yy}\left(\nu^{(1)}_{21} +\nu^{(1)}_{22}+\nu^{(2)}_{22}\right) - P^{(2)}_{yy} \nu^{(2)}_{12}}{\mathcal{D}},\label{eq:bar_P1xy}\\
	%%%
	\eta_2
	&=\frac{P^{(2)}_{yy}\left(\nu^{(1)}_{11} +\nu^{(1)}_{12} + \nu^{(2)}_{11}\right) - P^{(1)}_{yy} \nu^{(2)}_{21}}{\mathcal{D}}.\label{eq:bar_P2xy}
\end{align}
\end{subequations}
Here, the shear rate that appears in Eqs. \eqref{eq:eq5} and \eqref{eq:eq6} should be the same, which yields a condition to be satisfied by the partial temperatures, $T_1$ and $T_2$:
\begin{equation}
	\eta_1\left(\zeta_{21}+\zeta_{22}\right)
	= \eta_2\left(\zeta_{11}+\zeta_{12}\right).\label{eq:cond_T1_T2}
\end{equation}
When we fix the value of $T_1$, we can numerically obtain the value of $T_2$ from Eq. \eqref{eq:cond_T1_T2}.
Once the relation between $T_1$ and $T_2$ is determined from Eq. \eqref{eq:cond_T1_T2}, the shear rate is given by
\begin{equation}
	\dot\gamma^2 = \frac{3(\zeta_{11}+\zeta_{12}+\zeta_{21}+\zeta_{22})p}{2(\eta_1+\eta_2)}.
\end{equation}
We can also derive the shear viscosity, which is the sum of $\eta_i$:
\begin{equation}
	\eta = \eta_1 + \eta_2.
\end{equation}
%In the next section, we show results for various cases.

%%%%%%%%%%%%%%%%%%%%%%%%%%%%%%%%%%%%%%%%%
\section{Rheology}
Let us discuss the rheology of our system for some interesting cases. In Secs. \ref{sec:Mono} \NEU{and} \ref{sec:EqNo}, we consider special cases where the analysis becomes simple. Section \ref{sec:Gen} discusses a more general case.

%%%%%%%%%%%%%%%%%%%%%%%%%%%%%%%%%%%%%%%%%
\subsection{Monodisperse case ($x_1=1$, $x_2=0$)}
\label{sec:Mono}

We consider the most straightforward situation where only one species exists, i.e., $x_1=1$ and $x_2= 0$. In this case, the system is no longer a mixture. Therefore, there is no contribution from species $2$, thus, $\zeta_{12}=\zeta_{21}=\zeta_{22}=0$, $p_2=0$, $T_1=T$, and $p_1=p$. The shear rate and the shear viscosity are then
\begin{align}
  \dot\gamma^2 
  &= \frac{3}{2}\frac{\left(\nu^{(1)}_{11}+\nu^{(2)}_{11}\right)^2 \zeta_{11}}
    {\left(\nu^{(1)}_{11}+\nu^{(2)}_{11}\right)-\zeta_{11}},\\
  \eta 
  &= \frac{1}{\nu^{(1)}_{11}+\nu^{(2)}_{11}}\left(1-\frac{\zeta_{11}}{\nu^{(1)}_{11}+\nu^{(2)}_{11}}\right)p\,.
\end{align}
Apart from the contribution of the scattering angle, these expressions coincide with the corresponding expressions for gases whose particles interact via a square-well potential, published recently \cite{takada2018rheology} ($\zeta$ and $\nu$ in Ref. [\onlinecite{takada2018rheology}] correspond to $\zeta_{11}$ and $\nu^{(1)}_{11}+\nu^{(2)}_{11}$, respectively, in the current paper).

\begin{figure}[htbp]
	\includegraphics[width=\linewidth]{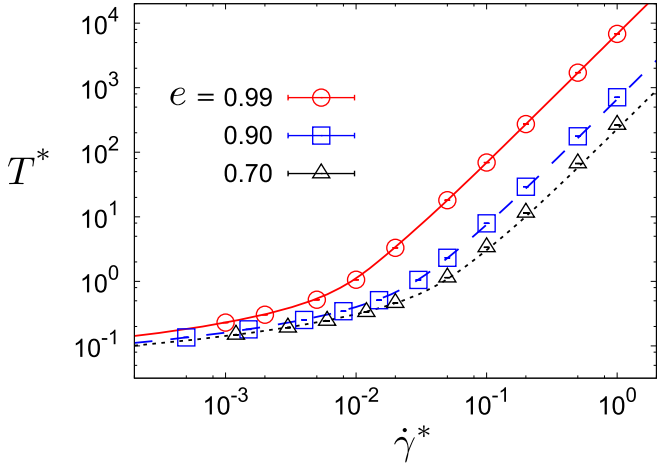}
	\caption{Granular temperature as a function of the shear rate for $e=0.99$ (solid line), $0.90$ (dashed line), and $0.70$ (dotted line), and $\lambda = 1.5$ in the monodisperse limit $x_1=1$.
	The open circles, squares, and triangles show the corresponding simulation results. 
    \NEU{
    For the explanation of the simulation data see Sec. \ref{sec:simulation}.}}
	\label{fig:T_shear_monodisperse}
\end{figure}

\begin{figure}[htbp]
	\includegraphics[width=\linewidth]{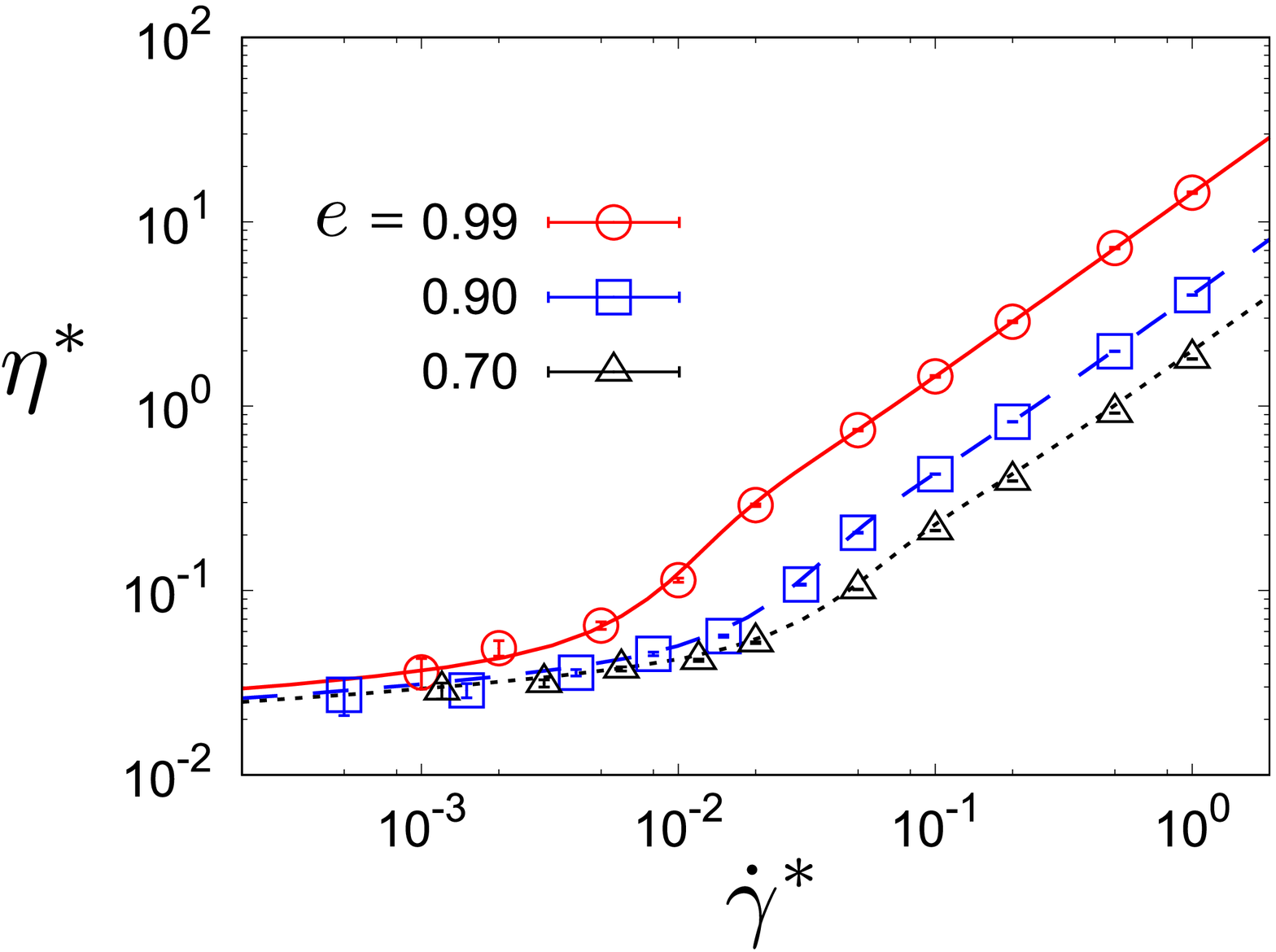}
	\caption{Shear viscosity as a function of the shear rate for $e=0.99$ (solid line), $0.90$ (dashed line), and $0.70$ (dotted line), and $\lambda = 1.5$ in the monodisperse limit $x_1=1$.
	The open circles, squares, and triangles show the corresponding  simulation results. 
    \NEU{
    For the explanation of the simulation data see Sec. \ref{sec:simulation}.}}
	\label{fig:eta_shear_monodisperse}
\end{figure}

Figures \ref{fig:T_shear_monodisperse} and \ref{fig:eta_shear_monodisperse}
%%%%%%%%%%%%%%%%%%%%%%%%%%%%%%%%%%%%%%%%%
%%%%%%%%%%%%%%%%%%%%%%%%%%%%%%%%%%%%%%%%%
%%%%%%%%%%%%%%%%%%%%%%%%%%%%%%%%%%%%%%%%%
show the shear rate \NEU{$\dot\gamma^*=\dot\gamma \sqrt{md^2/\varepsilon}$} dependences of the temperature \NEU{$T^* \equiv T/\varepsilon$} and the shear viscosity \NEU{$\eta^* \equiv \eta d^2/\sqrt{m\varepsilon}$}, respectively.
%Here, we introduce the following dimensionless quantities:
%\begin{align}
%	\dot\gamma^* = \dot\gamma \sqrt{\dfrac{md^2}{\varepsilon}},~T^* \equiv \dfrac{T}{\varepsilon},~\eta^* \equiv \eta \dfrac{d^2}{\sqrt{m\varepsilon}}.
 %   \label{eq:eta_T_gammmadot}
%\end{align}
For high shear rate, the temperature and the shear rate tend to $T\propto \dot\gamma^2$ and $\eta\propto \dot\gamma$, respectively. Here, the expressions approach Bagnold's results for hard-core gases \cite{santos2004inherent}:
\begin{align}
	T&=\frac{5\pi(2+e)}{432(1-e)(1+e)^2(3-e)^2}\frac{1}{\varphi^2}md^2\dot\gamma^2,\label{eq:T_Bagnold}\\
	\eta&=\frac{5(2+e)}{72(1+e)^2(3-e)^3}\sqrt{\frac{5(2+e)}{3(1-e)}}\frac{1}{\varphi}\frac{m}{d}\dot\gamma,\label{eq:eta_Bagnold}
\end{align}
where $\varphi=(\pi/6)nd^3$ is the packing fraction. 
This coincidence can be understood since for high shear rate, the shoulder of the potential is negligible, relative to the temperature, $\varepsilon/T\ll 1$. 
In contrast, for low shear rate, the temperature is nearly independent of the shear rate. Here, the shoulder prevents the particles to approach closely enough to enter the dissipative region. Therefore, the rate of inelastic collisions decreases in this regime, similar to the case in Refs. [\onlinecite{takada2017homogeneous},\onlinecite{takada2022transport}], and the temperature decays only weakly for  $T\lesssim \varepsilon$.
%Even if the energy injection due to shear is small, the energy dissipation rarely occurs for $T\lesssim \varepsilon$, which yields the steady temperature about $T\lesssim \varepsilon$.
\NEU{The behavior of the shear viscosity for low shear rate can be understood from the temperature as a function of the shear rate:
In a dilute hard-core granular system, the shear viscosity is proportional to the square root of the temperature, independent of the restitution coefficient \cite{chapman1970mathematical, brilliantov2004kinetic}.
On the other hand in the low-shear regime, the temperature is almost independent of the shear rate since the number of inelastic collisions decreases for $T\lesssim \varepsilon$. Therefore, here the shear viscosity is also nearly independent of the shear rate.
}
%Thus, shear viscosity depends on the shear rate for high shear rate.
%As previously mentioned, for low shear rate, the temperature is nearly independent of the shear rate.
%Therefore, shear viscosity is also nearly independent of the shear rate.

%%%%%%%%%%%%%%%%%%%%%%%%%%%%%%%%%%%%%%%%%
\subsection{Case for equal number and mechanical properties ($\NEU{x_1=x_2=1/2}$, $m_1=m_2$, $d_1=d_2$)}
\label{sec:EqNo}

Next, we consider the situation where the numbers of particles for both species are same, i.e., $x_1=x_2=1/2$.
Hereafter, we discuss only nearly elastic regime $e\lesssim 1$ because the kinetic theory for cohesive systems is limited in this regime \cite{takada2016kinetic,takada2018rheology}.
For this case, the system is symmetric with respect to the interchange of species $1\leftrightarrow2$.
This yields $\zeta_{11}=\zeta_{22}$, $\zeta_{12}=\zeta_{21}$, $\nu^{(i)}_{11}=\nu^{(i)}_{22}$, $\nu^{(i)}_{12}=\nu^{(i)}_{21}$ ($i=1,2$), and $T_1=T_2=T$, which means that Eq. \eqref{eq:cond_T1_T2} is satisfied.
Under these conditions, the shear rate and the shear viscosity become
\begin{align}
	\dot\gamma^2 
	&= \frac{3\left(\nu^{(1)}_{11}+\nu^{(1)}_{12}+\nu^{(2)}_{12}+\nu^{(2)}_{22}\right)^2 
	(\zeta_{11}+\zeta_{12})}{\nu^{(1)}_{11}+\nu^{(1)}_{12}+\nu^{(2)}_{12}+\nu^{(2)}_{22}-2(\zeta_{11}+\zeta_{12})},\\
	%%%
	\eta 
	&= \frac{\nu^{(1)}_{11}+\nu^{(1)}_{12}+\nu^{(2)}_{12}+\nu^{(2)}_{22}
	-2(\zeta_{11}+\zeta_{12})}{\left(\nu^{(1)}_{11}+\nu^{(1)}_{12}+\nu^{(2)}_{12}+\nu^{(2)}_{22}\right)^2}p.
\end{align}

\begin{figure}[htbp]
	\includegraphics[width=\linewidth]{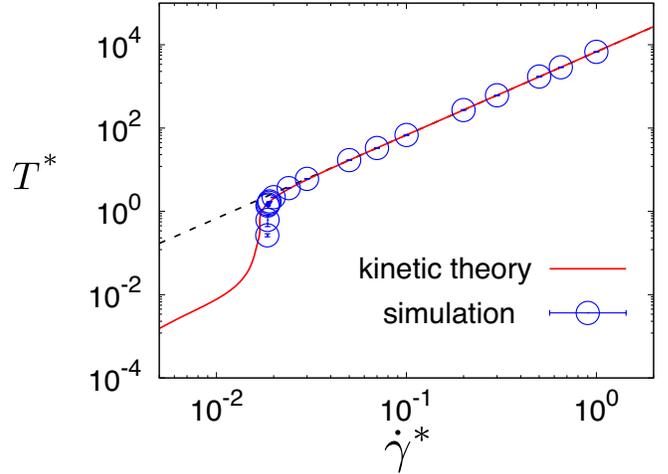}
	\caption{Temperature as a function of the shear rate for $e=0.99$, $\lambda = 1.5$, and $x_1=x_2=1/2$. The dashed line shows Bagnolds' scaling, Eq.  \eqref{eq:T_Bagnold}. \NEU{For the explanation of the simulation data see Sec. \ref{sec:simulation}.}}
	\label{fig:T_shear_0.50_0.50}
\end{figure}

\begin{figure}[htbp]
	\includegraphics[width=\linewidth]{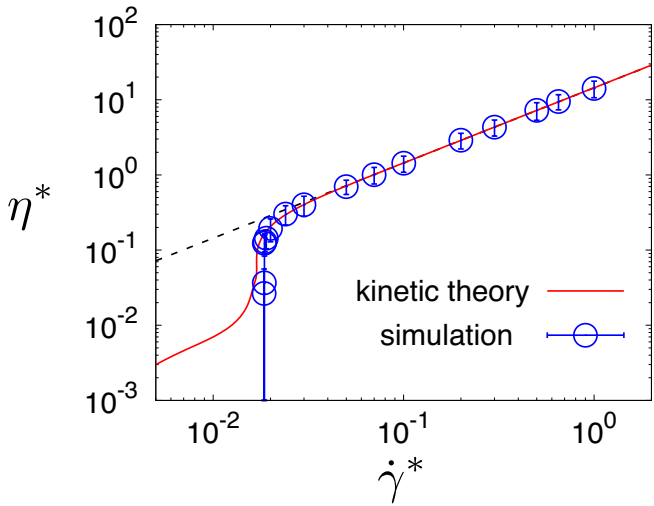}
	\caption{Shear viscosity as a function of the shear rate for $e=0.99$, $\lambda = 1.5$, and $x_1=x_2=1/2$. The dashed line shows Bagnolds' scaling, Eq. \eqref{eq:eta_Bagnold}. \NEU{For the explanation of the simulation data see Sec. \ref{sec:simulation}.}}
	\label{fig:eta_shear_0.50_0.50}
\end{figure}

Figures \ref{fig:T_shear_0.50_0.50} and \ref{fig:eta_shear_0.50_0.50} show the temperature and the shear viscosity as functions of the shear rate, respectively.
For $T\gtrsim \varepsilon$, the results are consistent with the Bagnolds' expressions.
On the other hand, the temperature drops almost discontinuously near $\dot\gamma^*\simeq 0.02$.
This critical shear rate corresponds to the point where Bagnolds' temperature, Eq. \eqref{eq:T_Bagnold}, becomes $T\simeq \varepsilon$.
This behavior is similar to that for the cohesive systems reported in Ref. \cite{takada2018rheology}, where the clustering processes is observed.

%%%%%%%%%%%%%%%%%%%%%%%%%%%%%%%%%%%%%%%%%
%%%%%%%%%%%%%%%%%%%%%%%%%%%%%%%%%%%%%%%%%
%%%%%%%%%%%%%%%%%%%%%%%%%%%%%%%%%%%%%%%%%

\subsection{General case -- unequal number and mechanical properties ($x_1\ge1/2$, $x_2\le 1/2$, $x_1+x_2=1$)}
\label{sec:Gen}

\begin{figure}[htbp]
    \includegraphics[width=\linewidth]{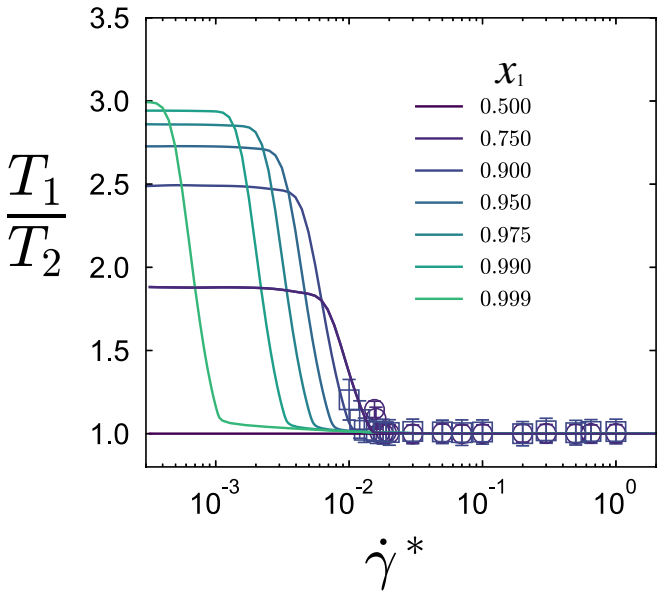}
    \caption{Temperature ratio $T_1/T_2$ as a function of the shear rate for $e=0.99$, $\lambda = 1.5$\NEU{, and various values of $x_1$. }
    \NEU{The simulation data for $x_1=0.750$ (open circles) and $0.900$ (open squares) are also plotted (see Sec. \ref{sec:simulation} for details).}}
    \label{fig:temp_ratio_0.75_0.25}
\end{figure}

In this subsection, let us consider the most general case.
For simplicity, we assume that $x_1\ge1/2$ and $x_2\le1/2$, \NEU{$x_1+x_2=1$} without loss of generality.
In this case, we need to solve Eq. \eqref{eq:cond_T1_T2} numerically to determine the two partial temperatures $T_1$ and $T_2$.
Figure \ref{fig:temp_ratio_0.75_0.25} shows the plot of the temperature ratio $T_1/T_2$ against the shear rate \NEU{for various values of $x_1$.}
The ratio tends to unity in the high shear limit because the potential depth is negligible compared to the temperature.
On the other hand, this ratio has a larger value when the shear rate decreases.
We have also found that the partial temperature of the majority is always larger than that of the minority.

\begin{figure}[htbp]
    \includegraphics[width=\linewidth]{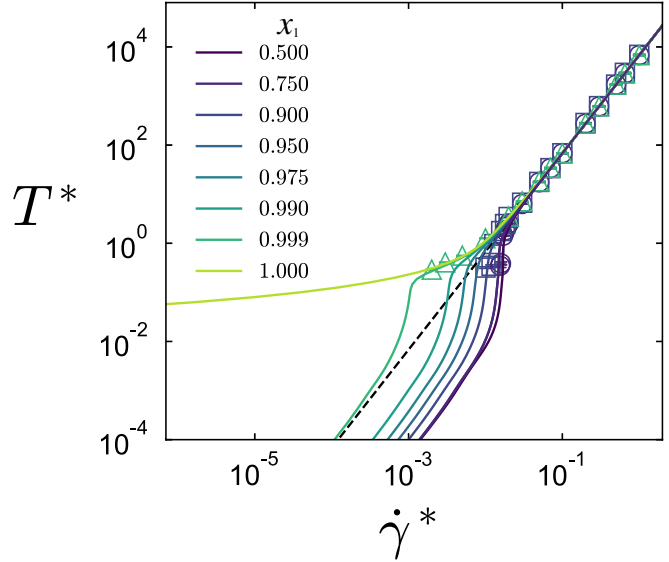}
    \caption{Temperature as a function of the shear rate for $e=0.99$, $\lambda = 1.5$, \NEU{and various values of $x_1$. }
    The dashed line shows  Bagnolds' scaling, Eq. \eqref{eq:T_Bagnold}. 
    \NEU{The simulation data for $x_1=0.750$ (open circles), $0.900$ (open squares), and $0.999$ (open triangles) are also plotted (see Sec. \ref{sec:simulation} for details).}}
    \label{fig:T_shear_0.75_0.25}
\end{figure}

\begin{figure}[htbp]
    \includegraphics[width=\linewidth]{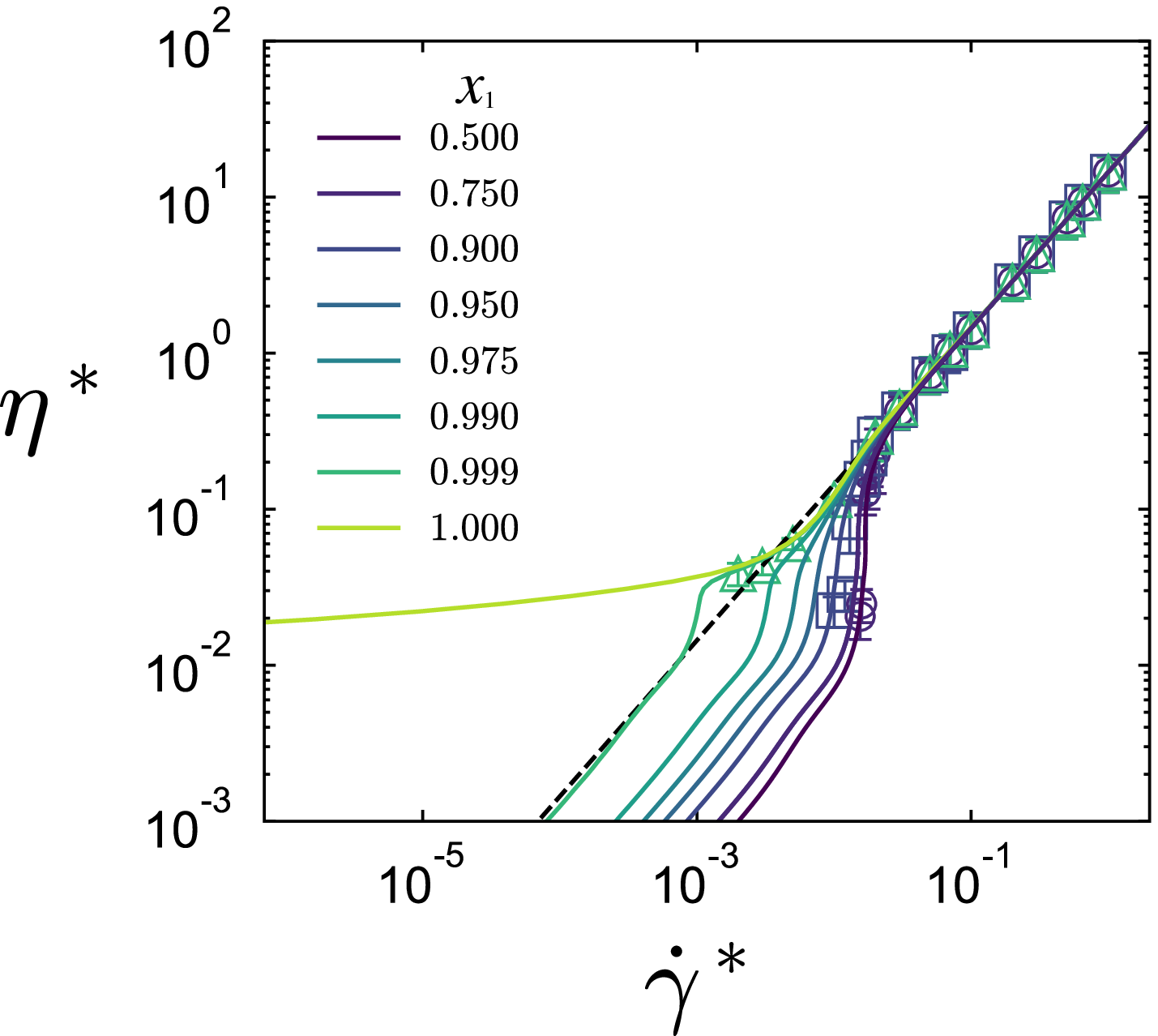}
    \caption{Shear viscosity as a function of the shear rate for $e=0.99$, $\lambda = 1.5$,\NEU{ and various values of $x_1$.}
    The dashed line shows Bagnolds' scaling, Eq. \eqref{eq:eta_Bagnold}. 
    \NEU{The simulation data for $x_1=0.750$ (open circles), $0.900$ (open squares), and $0.999$ (open triangles) are also plotted (see Sec. \ref{sec:simulation} for details)}}
    \label{fig:eta_shear_0.75_0.25}
\end{figure}

\NEU{Once the partial temperatures $T_1$ and $T_2$ are determined numerically, the total temperature $T$ is also given by $T=x_1 T_1+x_2 T_2$.
Then, we can evaluate the quantities as a function of $T_1$ and $T_2$.
Figures \ref{fig:T_shear_0.75_0.25} and \ref{fig:eta_shear_0.75_0.25} show the temperature and the shear viscosity as functions of the shear rate. For comparison, here we also plot the data for $x_1=1$ and $0.5$ explained in the previous subsections.
Similar to the cases $x_1=1$ and $0.5$, the temperature and the viscosity are consistent with Bagnolds’ expressions for the high shear velocity regime, disregarding the value of $x_1$.
On the other hand, the drops of the quantities appear at a certain critical shear rate, where this value decreases as the value of $x_1$ increases.}

%%%%%%%%%%%%%%%%%%%%%%%%%%%%%%%%%%%%%%%%%
%%%%%%%%%%%%%%%%%%%%%%%%%%%%%%%%%%%%%%%%%
%%%%%%%%%%%%%%%%%%%%%%%%%%%%%%%%%%%%%%%%%

%%%%%%%%%%%%%%%%%%%%%%%%%%%%%%%%%%%%%%%%%
\section{Simulation}\label{sec:simulation}
%%%%%%%%%%%%%%%%%%%%%%%%%%%%%%%%%%%%%%%%%
\begin{figure}[htbp]
	\includegraphics[width=\linewidth]{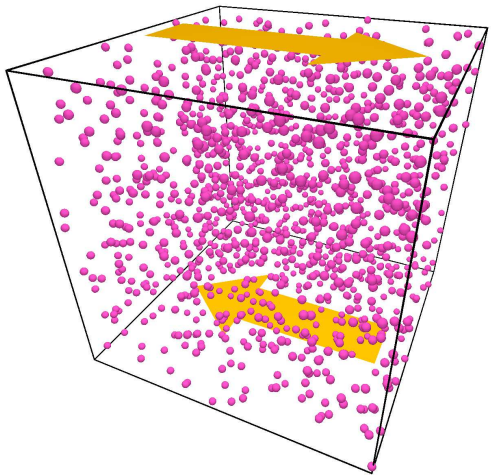}
	\caption{Snapshot of the system. The arrows indicate the directions of shear at $y=-L/2$ and $L/2$.}
	\label{fig:system}
\end{figure}
We also perform \NEU{event-driven} molecular dynamics simulations \NEU{using} DynamO \cite{bannerman2011dynamo} to address the validity of the kinetic theory explained in the previous section. We prepare $N=1372$ particles in the cubic box. They interact with each other via the square-shoulder (square-well) potential if two particles belong to the same (different) species. In this paper, we fix the packing fraction as $\varphi=0.01$ and accordingly the system size is $L=41.59d$. We divide the particles into two species, $N_1=\lfloor x_1 N\rfloor$ and $N_2=\lfloor x_2 N\rfloor$ particles belonging to species $1$ and $2$, respectively, where $\lfloor x \rfloor$ is the floor function. The shear is applied by Lees-Edwards boundary condition \cite{lees1972computer} in the $y$-direction. 
\NEU{We perform our simulation until $N_{\rm coll}=10^7$. For all cases the steady state of the system was achieved far earlier. Appendix \ref{sec:evolution} shows how the temperature and the shear viscosity converge with increasing number of collisions.}
A typical snapshot of the system is shown in Fig. \ref{fig:system}. As far as we have investigated, the system keeps uniform above the critical shear rate. In addition, the steady temperature and shear viscosity show good agreements with those obtained from the kinetic theory, which means that our theoretical treatment is valid in this regime.

\begin{figure}[htbp]
	\includegraphics[width=\linewidth]{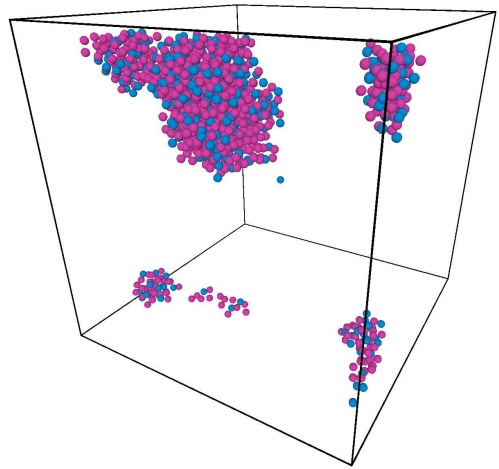}
	\caption{Snapshot of the system where clustering proceeds for $x_1=3/4$ and $x_2=1/4$. The color indicates the species of the particles.}
	\label{fig:clustering}
\end{figure}

As explained in the previous section, on the other hand, the clustering process proceeds as time goes on below the critical shear rate. After a long time, almost all particles are absorbed into larger clusters. The typical snapshot is shown in Fig. \ref{fig:clustering}. In this regime, the assumption of molecular chaos, which is important to develop the kinetic theory, is violated, and the treatment is no longer valid. It means that this regime is out of our theoretical treatment.

%%%%%%%%%%%%%%%%%%%%%%%%%%%%%%%%%%%%%%%%%
%%%%%%%%%%%%%%%%%%%%%%%%%%%%%%%%%%%%%%%%%

\section{Discussion and Summary}
\label{sec:discussion}
We studied the rheology of dilute granular gas mixtures of particles interacting via a square shoulder and well potential, using kinetic theory and numerical simulations. We have theoretically evaluated system's steady-state temperature, shear viscosity, and partial temperature ratio. The results converge to those by the Bagnold expressions in the high temperature (high shear) limit, in which we can regard the particle as hard-core gases. As the shear rate decreases to the critical value, the deviation from the Bagnoldian increases. We have also performed molecular dynamics simulations, and found that the simulation results are consistent with those from the kinetic theory above a critical shear rate, in which the system keeps uniform. On the other hand, the kinetic theory fails to reproduce the rheology of the system because the system is no longer uniform. In this regime, the clustering process proceeds as time goes on.

\NEU{Let us focus on the viscosity for the nearly elastic limit when the system is monodisperse.
  As shown in Fig. \ref{fig:eta_shear_monodisperse}, the slope of the viscosity for $e=0.99$ has a hump at $\dot\gamma^*\approx 10^{-2}$, which can be understood from the following argument:
  In Fig. \ref{fig:T_shear_monodisperse} we see that the shear rate $\dot\gamma^*\simeq 10^{-2}$  corresponds to $T\approx \varepsilon$. For $\dot\gamma^*\gtrsim 10^{-2}$ , colliding particles can overcome the square-shoulder of the potential and collide dissipatively. This yields nothing but the Bagnoldian expression of the viscosity. In contrast, for $\dot\gamma^*\lesssim 10^{-2}$, an increasing fraction of collisions occur elastically, due to the reflection at $r_{ij}=\lambda d_{ij}$.
  % This means that the number of elastic collision increases.
  We can introduce the effective restitution coefficient $e_\mathrm{eff}$, which is the mean restitution coefficient when we consider both elastic and inelastic collisions, thus, $e< e_\mathrm{eff}< 1$.
By substituting $e_\mathrm{eff}$ into Eq. \eqref{eq:eta_Bagnold}, we can easily show that the coefficient of the right-hand side becomes larger.
This is why we see a hump in Fig. \ref{fig:eta_shear_monodisperse}.}

For the numerical work of the current paper, we have assumed that all the particles have the same size and diameter. However, many systems of practical interest are disperse, e.g., in the context of planetary science. The study of such systems will be left for our future work.

It is also known that charges can be transferred from one particle to another during collisions. We believe that such processes can be modeled by the kinetic theory employing Smoluchowski-like equations \cite{brilliantov2015size,osinsky2022anomalous}, which will also be subject of our future work.

\section*{ACKNOWLEDGMENTS}
The numerical computation was partially carried out at the Yukawa Institute Computer Facility.  K.Y. was supported by the Grant-in-Aid for Japan Society for Promotion of Science, JSPS Research Fellow (Grant No. 21J13720) and Hosokawa Powder Technology Foundation (No. HPTF20506).  S.T. was supported by Scientific Grant-in-Aid of Japan Society for the Promotion of Science, KAKENHI (Grants No. \NEU{20K14428} and No. 21H01006). K.K. was supported by Scientific Grant-in-Aid of Japan Society for the Promotion of Science, KAKENHI (Grant No. \NEU{17K18812}). This work was supported by the Interdisciplinary Center for Nanostructured Films (IZNF), the Competence Unit for Scientific Computing (CSC), and the Interdisciplinary Center for Functional Particle Systems (FPS) at Friedrich-Alexander-Universit\"at Erlangen-N\"urnberg.

\NEU{\section*{DATA AVAILABILITY}
The data relating to the findings of this study are available from the corresponding author upon reasonable request.
}

\appendix
\begin{widetext}
\section{Detailed derivation of the second moment of the collision integral $\Lambda_{\alpha\beta}^{(ij)}$}\label{sec:Lambda}
In this appendix, we present the detailed derivation of the moment $\Lambda^{(ij)}_{\alpha\beta}$.
For this purpose, we introduce the dimensionless velocities
\begin{equation}
	\displaystyle \bm{g} =\frac{\bm{V}_1 - \bm{V}_2}{v_{\rm T}},\quad
	\displaystyle \bm{G} = \frac{\mu_{ij}\bm{V}_1 + \mu_{ji}\bm{V}_2}{v_{\rm T}} +\frac{\mu_{ji}\vartheta_i-\mu_{ij}\vartheta_j}{\vartheta_i+\vartheta_j}\bm{g}.
	\label{eq:G_g}
\end{equation}
Using Eqs. \eqref{eq:relation_pre_post_vel} and \eqref{eq:G_g}, the relationship between the pre- and post-collisional velocities as
\begin{equation}
	m_i v_{1,\alpha}^\prime v_{1,\beta}^\prime - m_i v_{1,\alpha} v_{1,\beta}
	= -m_i \mu_{ji}A_{ij}v_{\rm T}^2 (\bm{g}\cdot \widehat{\bm{\sigma}})
		\left[G_\alpha \widehat{\sigma}_\beta + G_\beta \widehat{\sigma}_\alpha
			+\frac{\vartheta_j}{\vartheta_i+\vartheta_j}(g_\alpha \widehat{\sigma}_\beta + g_\beta \widehat{\sigma}_\alpha)
			-\mu_{ji}A_{ij}(\bm{g}\cdot \widehat{\bm{\sigma}})\widehat{\sigma}_\alpha \widehat{\sigma}_\beta\right].
	\label{eq:mv1pv1p-mv1v1}
\end{equation}
Similarly, we can rewrite $f_i(\bm{V}_1)f_j(\bm{V}_2)$ as
\begin{align}
	f_i\left(\bm{V}_1\right) f_j\left(\bm{V}_2\right)
	&= n_i n_j  v_{\rm T}^{-3} \left(\vartheta_i \vartheta_j\right)^{3/2}
	\pi^{-3} \exp\left[-(\vartheta_i + \vartheta_j)G^2 -\frac{\vartheta_i \vartheta_j}{\vartheta_i + \vartheta_j}g^2\right]\nonumber\\
	&\hspace{1em}\times
	\left[1+\vartheta_i \Pi^{(i)}_{\alpha\beta}
		\left(G_\alpha+\frac{\vartheta_j}{\vartheta_i + \vartheta_j}g_\alpha\right)
		\left(G_\beta+\frac{\vartheta_j}{\vartheta_i + \vartheta_j}g_\beta\right)
		+\vartheta_j \Pi^{(j)}_{\alpha\beta}
		\left(G_\alpha-\frac{\vartheta_i}{\vartheta_i + \vartheta_j}g_\alpha\right)
		\left(G_\beta-\frac{\vartheta_i}{\vartheta_i + \vartheta_j}g_\beta\right)\right].
	\label{eq:fi_fj_linear}
\end{align}
Then, from Eqs. \eqref{eq:mv1pv1p-mv1v1} and \eqref{eq:fi_fj_linear}, Eq. \eqref{eq:Lambda_ij} is rewritten as
\begin{align}
	\Lambda^{(ij)}_{\alpha\beta}
	&= m_i \mu_{ji} n_i n_j d_{ij}^{2}(\vartheta_i\vartheta_j)^{3/2} v_{\rm T}^3 \widetilde{\Lambda}^{(ij)}_{\alpha\beta},
	\label{eq:Lambda_dimensionless}
\end{align}
with the linear collisional moment
\begin{align}
	\widetilde{\Lambda}^{(ij)}_{\alpha\beta}
	&\equiv \frac{1}{\pi^3}
	\int d\bm{G} \int d\bm{g} \int d\widehat{\bm{\sigma}} A_{ij}
	\sigma_{\rm s}^* |\widehat{\bm{\sigma}}\cdot \bm{g}| 
	(\widehat{\bm{\sigma}}\cdot \bm{g})
	\exp\left[-(\vartheta_i + \vartheta_j)G^2 -\frac{\vartheta_i \vartheta_j}{\vartheta_i + \vartheta_j}g^2\right]\nonumber\\
	&\hspace{1em}\times
	\left[G_\alpha \widehat{\sigma}_\beta + G_\beta \widehat{\sigma}_\alpha
			+\frac{\vartheta_j}{\vartheta_i+\vartheta_j}(g_\alpha \widehat{\sigma}_\beta + g_\beta \widehat{\sigma}_\alpha)
			-\mu_{ji}A_{ij}(\bm{g}\cdot \widehat{\bm{\sigma}})\widehat{\sigma}_\alpha \widehat{\sigma}_\beta\right]\nonumber\\
	&\hspace{1em}\times
	\left[1+\vartheta_i \Pi^{(i)}_{\gamma\delta}
		\left(G_\gamma+\frac{\vartheta_j}{\vartheta_i + \vartheta_j}g_\gamma\right)
		\left(G_\delta+\frac{\vartheta_j}{\vartheta_i + \vartheta_j}g_\delta\right)
		+\vartheta_j \Pi^{(j)}_{\gamma\delta}
		\left(G_\gamma-\frac{\vartheta_i}{\vartheta_i + \vartheta_j}g_\gamma\right)
		\left(G_\delta-\frac{\vartheta_i}{\vartheta_i + \vartheta_j}g_\delta\right)\right].
	\label{eq:tilde_Lambda}
\end{align}
Here, we have introduced the dimensionless collision cross section $\sigma_{\rm s}^*\equiv \sigma_{\rm s}/d_{ij}^2$.
For further calculation, it is convenient to introduce $\widetilde{I}_{ij}^{(\ell)}(\bm{g}, \widehat{\bm{\sigma}})$, $\widetilde{I}_{ij,\alpha}^{(\ell)}(\bm{g}, \widehat{\bm{\sigma}})$, and $\widehat{I}_{ij,\alpha}^{(\ell)}(\bm{g}, \widehat{\bm{\sigma}})$ \cite{takada2020enskog} as
\begin{subequations}
\begin{align}
	\begin{Bmatrix} \widetilde{I}_{ij}^{(\ell)}(\bm{g}, \widehat{\bm{\sigma}}) \\ \widetilde{I}_{ij,\alpha}^{(\ell)}(\bm{g}, \widehat{\bm{\sigma}}) \end{Bmatrix}
	&\equiv \frac{1}{\pi^3} \int d\bm{G}
		\sigma_{\rm s}^* |\widehat{\bm{\sigma}}\cdot \bm{g}| 
		(\widehat{\bm{\sigma}}\cdot \bm{g})^{\ell-1}
		\begin{Bmatrix} 1 \\ g_\alpha \end{Bmatrix}
		\exp\left[-(\vartheta_i + \vartheta_j)G^2 -\frac{\vartheta_i \vartheta_j}{\vartheta_i + \vartheta_j}g^2\right]\nonumber\\
	&\hspace{1em}\times
	\left[1+\vartheta_i \Pi^{(i)}_{\gamma\delta}
		\left(G_\gamma+\frac{\vartheta_j}{\vartheta_i + \vartheta_j}g_\gamma\right)
		\left(G_\delta+\frac{\vartheta_j}{\vartheta_i + \vartheta_j}g_\delta\right)
		+\vartheta_j \Pi^{(j)}_{\gamma\delta}
		\left(G_\gamma-\frac{\vartheta_i}{\vartheta_i + \vartheta_j}g_\gamma\right)
		\left(G_\delta-\frac{\vartheta_i}{\vartheta_i + \vartheta_j}g_\delta\right)\right]\nonumber\\
	%%%
	&= \frac{1}{\pi^{3/2}(\vartheta_i + \vartheta_j)^{3/2}} 
		\sigma_{\rm s}^* |\widehat{\bm{\sigma}}\cdot \bm{g}| 
		(\widehat{\bm{\sigma}}\cdot \bm{g})^{\ell-1}
		\begin{Bmatrix} 1 \\ g_\alpha \end{Bmatrix}
		\exp\left(-\frac{\vartheta_i \vartheta_j}{\vartheta_i + \vartheta_j}g^2\right)\nonumber\\
		&\hspace{1em}\times\left[1 + \frac{\vartheta_i \vartheta_j^2}{(\vartheta_i+\vartheta_j)^2}g_\gamma g_\delta \Pi^{(i)}_{\gamma\delta}
		+ \frac{\vartheta_i^2 \vartheta_j}{(\vartheta_i+\vartheta_j)^2}g_\gamma g_\delta \Pi^{(j)}_{\gamma\delta}\right],
	\label{eq:tilde_I}\\
	%%%
	%%%
	\widehat{I}_{ij,\alpha}^{(\ell)}(\bm{g}, \widehat{\bm{\sigma}}) 
	&\equiv \frac{1}{\pi^3} \int d\bm{G}
		\sigma_{\rm s}^* |\widehat{\bm{\sigma}}\cdot \bm{g}| 
		(\widehat{\bm{\sigma}}\cdot \bm{g})^{\ell-1}
		G_\alpha
		\exp\left[-(\vartheta_i + \vartheta_j)G^2 -\frac{\vartheta_i \vartheta_j}{\vartheta_i + \vartheta_j}g^2\right]\nonumber\\
	&\hspace{1em}\times
	\left[1+\vartheta_i \Pi^{(i)}_{\gamma\delta}
		\left(G_\gamma+\frac{\vartheta_j}{\vartheta_i + \vartheta_j}g_\gamma\right)
		\left(G_\delta+\frac{\vartheta_j}{\vartheta_i + \vartheta_j}g_\delta\right)
		+\vartheta_j \Pi^{(j)}_{\gamma\delta}
		\left(G_\gamma-\frac{\vartheta_i}{\vartheta_i + \vartheta_j}g_\gamma\right)
		\left(G_\delta-\frac{\vartheta_i}{\vartheta_i + \vartheta_j}g_\delta\right)\right]\nonumber\\
	%%%
	&= \frac{1}{\pi^{3/2}(\vartheta_i + \vartheta_j)^{3/2}} 
		\sigma_{\rm s}^* |\widehat{\bm{\sigma}}\cdot \bm{g}| 
		(\widehat{\bm{\sigma}}\cdot \bm{g})^{\ell-1}
		\exp\left(-\frac{\vartheta_i \vartheta_j}{\vartheta_i + \vartheta_j}g^2\right)
		\frac{\vartheta_i \vartheta_j}{(\vartheta_i + \vartheta_j)^2} g_\gamma 
		\left(\Pi^{(i)}_{\alpha\gamma} - \Pi^{(j)}_{\alpha\gamma}\right).
	\label{eq:hat_I}
\end{align}
\end{subequations}
Using Eqs. \eqref{eq:tilde_I} and \eqref{eq:hat_I}, Eq. \eqref{eq:tilde_Lambda} is rewritten as
\begin{align}
	\widetilde{\Lambda}^{(ij)}_{\alpha\beta}
	&=
	\int d\bm{g}\int d\widehat{\bm{\sigma}}A_{ij}
	\left\{\widehat{I}^{(2)}_{ij, \alpha}(\bm{g}, \widehat{\bm{\sigma}}) \widehat{\sigma}_\beta 
		+ \widehat{I}^{(2)}_{ij, \beta}(\bm{g}, \widehat{\bm{\sigma}}) \widehat{\sigma}_\alpha
		+ \frac{\vartheta_j}{\vartheta_i+\vartheta_j}\left[
			\widetilde{I}^{(2)}_{ij, \alpha}(\bm{g}, \widehat{\bm{\sigma}}) \widehat{\sigma}_\beta 
			+ \widetilde{I}^{(2)}_{ij, \beta}(\bm{g}, \widehat{\bm{\sigma}}) \widehat{\sigma}_\alpha
		\right]
		- \mu_{ji}A_{ij}\widetilde{I}^{(3)}_{ij}(\bm{g}, \widehat{\bm{\sigma}}) \widehat{\sigma}_\alpha \widehat{\sigma}_\beta
	\right\}.
	\label{eq:tilde_Lambda_2}
\end{align}
Now, let us use the following identities \cite{chapman1970mathematical}:
\begin{subequations}
\begin{align}
	\int d\widehat{\bm{\sigma}} \sigma_{\rm s}^* |\widehat{\bm{\sigma}}\cdot\bm{g}|
	(\widehat{\bm{\sigma}}\cdot\bm{g})\widehat{\sigma}_\alpha
	&= 2\pi \int_0^\infty db_{ij}^* b_{ij}^* g g_\alpha \sin^2\frac{\chi_{ij}}{2}.\label{eq:identity1}\\
	%%%
	\int d\widehat{\bm{\sigma}} \sigma_{\rm s}^* |\widehat{\bm{\sigma}}\cdot\bm{g}|
	(\widehat{\bm{\sigma}}\cdot\bm{g})^2\widehat{\sigma}_\alpha \widehat{\sigma}_\beta
	&= \pi \int_0^\infty db_{ij}^* b_{ij}^* g \sin^2\frac{\chi_{ij}}{2}
	\left[g^2 \cos^2\frac{\chi_{ij}}{2}\delta_{\alpha\beta} 
	+ \left(2\sin^2\frac{\chi_{ij}}{2}-\cos^2\frac{\chi_{ij}}{2}\right)g_\alpha g_\beta\right].
	\label{eq:identity2}
\end{align}
\end{subequations}
Then, one gets
\begin{subequations}
\begin{align}
	\int d\bm{g}\int d\widehat{\bm{\sigma}}A_{ij}
	\widetilde{I}^{(2)}_{ij, \alpha}(\bm{g}, \widehat{\bm{\sigma}}) \widehat{\sigma}_\beta 
	%%%
	&= \frac{1}{\pi^{3/2}(\vartheta_i + \vartheta_j)^{3/2}} \frac{8\pi^2}{3}
		\int_0^\infty dg \int_0^\infty db_{ij}^* A_{ij} b_{ij}^* g^5\sin^2\frac{\chi_{ij}}{2}
		\exp\left(-\frac{\vartheta_i \vartheta_j}{\vartheta_i + \vartheta_j}g^2\right)\nonumber\\
		&\hspace{1em}\times\left[\delta_{\alpha\beta}
		+ \frac{2}{5}\frac{\vartheta_i \vartheta_j^2}{(\vartheta_i+\vartheta_j)^2}g^2 \Pi^{(i)}_{\alpha\beta}
		+ \frac{2}{5}\frac{\vartheta_i^2 \vartheta_j}{(\vartheta_i+\vartheta_j)^2}g^2 \Pi^{(j)}_{\alpha\beta}\right],\\	
	\label{eq:tilde_I_2}
	%%%
	%%%
	\int d\bm{g}\int d\widehat{\bm{\sigma}}A_{ij}^2
	\widetilde{I}^{(3)}_{ij}(\bm{g}, \widehat{\bm{\sigma}}) \widehat{\sigma}_\alpha \widehat{\sigma}_\beta
	%%%
	&= \frac{1}{\pi^{3/2}(\vartheta_i + \vartheta_j)^{3/2}} \frac{8\pi^2}{3} 
		\int_0^\infty dg \int_0^\infty db_{ij}^* A_{ij}^2 b_{ij}^* g^5 \sin^2\frac{\chi_{ij}}{2}
		\exp\left(-\frac{\vartheta_i \vartheta_j}{\vartheta_i + \vartheta_j}g^2\right)\nonumber\\
		&\hspace{1em}\times\left\{\delta_{\alpha\beta} + \frac{2}{5}g^2 \left(1-\frac{3}{2}\cos^2\frac{\chi_{ij}}{2}\right)
		\left[\frac{\vartheta_i \vartheta_j^2}{(\vartheta_i+\vartheta_j)^2} \Pi^{(i)}_{\alpha\beta}
		+ \frac{\vartheta_i^2 \vartheta_j}{(\vartheta_i+\vartheta_j)^2} \Pi^{(j)}_{\alpha\beta}\right]\right\},\\
	%%%
	%%%
	\int d\bm{g}\int d\widehat{\bm{\sigma}}A_{ij}
	\widehat{I}^{(2)}_{ij, \alpha}(\bm{g}, \widehat{\bm{\sigma}}) \widehat{\sigma}_\beta
	%%%
	&= \frac{1}{\pi^{3/2}(\vartheta_i + \vartheta_j)^{3/2}} \frac{8\pi^2}{3} 
		\int_0^\infty dg \int_0^\infty db_{ij}^* A_{ij} b_{ij}^* g^5 \sin^2\frac{\chi_{ij}}{2}
		\exp\left(-\frac{\vartheta_i \vartheta_j}{\vartheta_i + \vartheta_j}g^2\right)\nonumber\\
		&\hspace{1em}\times\frac{\vartheta_i \vartheta_j}{(\vartheta_i + \vartheta_j)^2}  
		\left(\Pi^{(i)}_{\alpha\beta} - \Pi^{(j)}_{\alpha\beta}\right).
	\label{eq:hat_I_2}
\end{align}
\end{subequations}
Substituting Eqs. \eqref{eq:tilde_I_2}--\eqref{eq:hat_I_2} into Eq. \eqref{eq:tilde_Lambda_2}, the dimensionless form $\widetilde{\Lambda}^{(ij)}_{\alpha\beta}$ becomes
\begin{align}
	\widetilde{\Lambda}^{(ij)}_{\alpha\beta}
	&=
	\frac{8\sqrt{\pi}}{3(\vartheta_i + \vartheta_j)^{3/2}}  
		\int_0^\infty dg \int_0^\infty db_{ij}^* A_{ij} b_{ij}^* g^5 \sin^2\frac{\chi_{ij}}{2}
		\exp\left(-\frac{\vartheta_i \vartheta_j}{\vartheta_i + \vartheta_j}g^2\right)\nonumber\\
		&\hspace{1em}\times\left(
			\left[\frac{2\vartheta_j}{\vartheta_i+\vartheta_j}-\mu_{ji}A_{ij}\right]\delta_{\alpha\beta}
			+\frac{2\vartheta_i\vartheta_j}{\vartheta_i+\vartheta_j}
			\left\{1+\frac{\vartheta_j}{5}g^2\left[\frac{2\vartheta_j}{\vartheta_i+\vartheta_j} 
				-\mu_{ji}A_{ij}\left(1-\frac{3}{2}\cos^2\frac{\chi_{ij}}{2}\right)\right]\right\}\Pi^{(i)}_{\alpha\beta}\right.\nonumber\\
		&\hspace{3em}\left.
			-\left\{1-\frac{\vartheta_j}{5}g^2\left[\frac{2\vartheta_i}{\vartheta_i+\vartheta_j} 
				-\mu_{ji}A_{ij}\left(1-\frac{3}{2}\cos^2\frac{\chi_{ij}}{2}\right)\right]\right\}\Pi^{(j)}_{\alpha\beta}
		\right).
	\label{eq:tilde_Lambda_3}
\end{align}
or equivalently, the dimensional form $\Lambda^{(ij)}_{\alpha\beta}$ becomes
\begin{align}
	\Lambda^{(ij)}_{\alpha\beta}
	&=
	\frac{8\sqrt{\pi}}{3} 
		m_i \mu_{ji} n_i n_j d_{ij}^{2}\left(\frac{\vartheta_i\vartheta_j}{\vartheta_i + \vartheta_j}\right)^{3/2} v_{\rm T}^3
		\int_0^\infty dg \int_0^\infty db_{ij}^* A_{ij} b_{ij}^* g^5 \sin^2\frac{\chi_{ij}}{2}
		\exp\left(-\frac{\vartheta_i \vartheta_j}{\vartheta_i + \vartheta_j}g^2\right)\nonumber\\
		&\hspace{1em}\times\left(
			\left[\frac{2\vartheta_j}{\vartheta_i+\vartheta_j}-\mu_{ji}A_{ij}\right]\delta_{\alpha\beta}
			+\frac{2\vartheta_i\vartheta_j}{\vartheta_i+\vartheta_j}
			\left\{1+\frac{\vartheta_j}{5}g^2\left[\frac{2\vartheta_j}{\vartheta_i+\vartheta_j} 
				-\mu_{ji}A_{ij}\left(1-\frac{3}{2}\cos^2\frac{\chi_{ij}}{2}\right)\right]\right\}\Pi^{(i)}_{\alpha\beta}\right.\nonumber\\
		&\hspace{3em}\left.
			-\left\{1-\frac{\vartheta_j}{5}g^2\left[\frac{2\vartheta_i}{\vartheta_i+\vartheta_j} 
				-\mu_{ji}A_{ij}\left(1-\frac{3}{2}\cos^2\frac{\chi_{ij}}{2}\right)\right]\right\}\Pi^{(j)}_{\alpha\beta}
		\right),
	\label{eq:tilde_Lambda_4}
\end{align}
which is equivalent to Eq. \eqref{eq:Lambda_ij} with Eqs. \eqref{eq:def_zeta_nu}.
\end{widetext}

\NEU{\section{Evolution of temperature and shear viscosity}\label{sec:evolution}
In this appendix, we discuss the evolution of the temperature and the shear viscosity obtained from the simulation.
\begin{figure}[htbp]
		\includegraphics[width=0.9\linewidth]{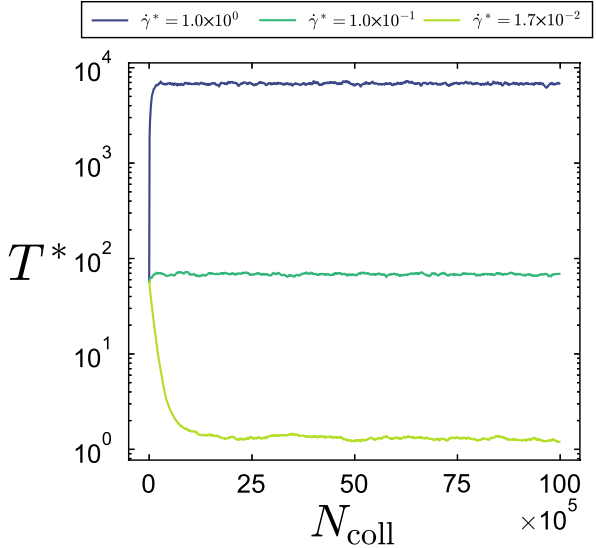}
		\caption{\NEU{Evolution of the temperature against the number of collisions $N_{\rm coll}$.}}
		\label{fig:evolution_T}
\end{figure}
\begin{figure}[htbp]
		\includegraphics[width=0.9\linewidth]{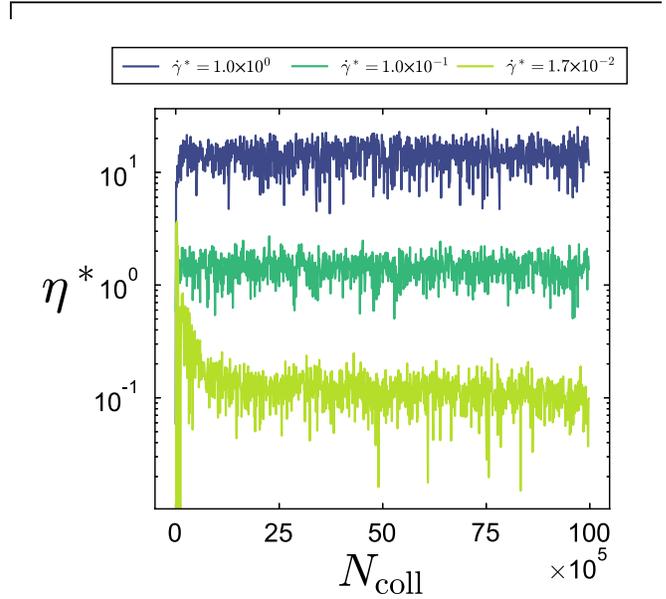}
		\caption{\NEU{Evolution of the shear viscosity against the number of collisions $N_{\rm coll}$.}}
		\label{fig:evolution_eta}
\end{figure}
Figures \ref{fig:evolution_T} and \ref{fig:evolution_eta} show their evolution for $\dot{\gamma}^* = 1.0\times 10^{0}$, $1.0\times 10^{-1}$, and $1.7\times 10^{-2}$ for the case of equal fractions, i.e., $x_1=x_2=1/2$, $m_1=m_2$, $d_1=d_2$. 
For all cases, the quantities converge to constants after $N_{\rm coll}\sim 25\times10^{5}$ collisions in total, corresponding to ca. $2000$ collisions per particle. 
In the main text, we use the time-averaged quantities obtained from these time series.
This result is almost independent of $x_1$ and $x_2$.
%It is noted that the behavior is almost the same even for $x_1\neq x_2$.
}

%\bibliography{aipsamp}
\bibliography{Ref_charged}

\end{document}